\documentclass{ifacconf}

\usepackage{graphicx}      
\usepackage{natbib}        

\usepackage{graphics} 
\usepackage{epsfig} 
\usepackage{mathptmx} 
\usepackage{times} 
\usepackage{amsmath} 
\usepackage{amssymb}  

\usepackage{color}
\usepackage{float}
\usepackage{graphics}

\usepackage{calrsfs}
\DeclareMathAlphabet{\pazocal}{OMS}{zplm}{m}{n}

\usepackage{algorithm,algcompatible,amsmath}
\algnewcommand\INPUT{\item[\textbf{Input:}]}%
\algnewcommand\OUTPUT{\item[\textbf{Output:}]}%

\newtheorem{assumption}{Assumption}[section]

\begin{document}
\begin{frontmatter}

\title{Distributed Model Predictive Control with Asymmetric Adaptive Terminal Sets for the Regulation  of Large-scale Systems \thanksref{footnoteinfo}} 

\thanks[footnoteinfo]{Research supported by the Swiss Innovation Agency Innosuisse under the Swiss Competence Center for Energy Research SCCER FEEB$\&$D.}

\author[First]{Ahmed Aboudonia}, 
\author[Second]{Annika Eichler}, 
\author[First]{John Lygeros}

\address[First]{Automatic Control Laboratory, Department of Electrical Engineering and Information Technology, ETH Zurich, Switzerland (e-mail: $\{$ahmedab, jlygeros$\}$@control.ee.ethz.ch)}
\address[Second]{Deutsches Elektronen-Synchroton DESY, 22607
	Hamburg, Germany (e-mail: annika.eichler@desy.de)}

\begin{abstract}                
In this paper, a novel distributed model predictive
control (MPC) scheme with asymmetric adaptive terminal sets
is developed for the regulation of large-scale systems with
a distributed structure. Similar to typical MPC schemes,
a structured Lyapunov matrix and a distributed terminal
controller, respecting the distributed structure of the system,
are computed offline. However, in this scheme, a distributed
positively invariant terminal set is computed online and updated
at each time instant taking into consideration the current
state of the system. In particular, we consider ellipsoidal terminal sets as they are easy to compute for large-scale systems. The size and the center of these terminal sets, together with
the predicted state and input trajectories, are considered as
decision variables in the online phase. Determining the terminal
set center online is found to be useful specifically in the
presence of asymmetric constraints. Finally, a relaxation of
the resulting online optimal control problem is provided.
The efficacy of the proposed scheme is illustrated in simulation
by comparing it to a recent distributed MPC scheme with adaptive
terminal sets.
\end{abstract}

\begin{keyword}
Predictive Control, Invariance, Distributed Control, Large-scale Systems, Adaptive Control 
\end{keyword}

\end{frontmatter}

\section{Introduction}

Thanks to its flexibility, versatility and strong theoretical properties \citep{kouvaritakis2016model}, Model Predictive Control (MPC) has been used over the past years in many practical applications such as robotics \citep{klanvcar2007tracking}, energy management \citep{prodan2014model,scherer2014efficient,zeng2015parallel}, and systems biology \citep{hovorka2004nonlinear} to name a few. Besides, many MPC variants have been developed including, but not limited to, robust MPC \citep{bemporad1999robust}, stochastic MPC \citep{mesbah2016stochastic} and economic MPC \citep{ellis2014tutorial}.

MPC is typically designed in a centralized fashion with one optimization problem solved for the whole controlled plant. For large-scale distributed systems such as power systems and water networks, centralized MPC may lead to communication and computational complications \citep{christofides2013distributed}. To overcome these difficulties, distributed MPC techniques have been developed to decompose the large-scale system into several smaller subsystems and design a local controller for each. 

Due to the increasing interest in MPC in different applications, various efforts have been devoted to ensure the closed loop stability of plants controlled using MPC \citep{mayne2000constrained}. A well-known method for ensuring asymptotic stability and recursive feasibility is the addition of a terminal cost and/or a terminal constraint. This method has been extensively used for centralized MPC, see, for example, \cite{keerthi1988optimal,rawlings1993stability,sznaier1987suboptimal}. It has also been extended to distributed MPC, by using a quadratic terminal cost and an ellipsoidal terminal set \citep{conte2012distributed,conte2016distributed}. 

In most cases, the terminal set is computed without taking the system's current state into account, possibly resulting in small regions of attraction. Recently, a novel distributed MPC scheme with adaptive terminal sets was proposed in \cite{darivianakis2019distributed}. In this scheme, an ellipsoidal terminal set is determined and updated online based on the current state of the system, yielding a larger domain of attraction.

In this work, a novel distributed MPC with ellipsoidal asymmetric adaptive terminal sets is developed for regulating constrained large-scale linear time-invariant systems. One advantage of this approach over the one introduced in  \cite{darivianakis2019distributed} is that the terminal set is not centered at the origin. Instead, the center of the terminal set, together with its size, are assumed to be decision variables to be determined online. The online computation of the terminal set center results generally in enlarging the feasible region. The terminal set invariance and constraint satisfaction are guaranteed through the addition of extra constraints formulated as linear matrix inequalities (LMIs) in the online optimization problem. A relaxation of the derived LMIs is performed by directly using the linear state and input constraints instead of the quadratic ones in \cite{darivianakis2019distributed}. This relaxation is found to be very useful in the case of asymmetric state and input constraints. The effectiveness of this approach is evaluated by means of a simulation example.

In Section II, the distributed MPC problem is formulated. In Section III, the offline phase in which the terminal cost and terminal controller are computed is presented for the sake of completeness. Moreover, the online phase of the distributed MPC scheme with asymmetric adaptive terminal sets, which is the main contribution of this work, is presented. In Section IV, a numerical simulation illustrates the efficacy of this scheme. Finally, concluding remarks are mentioned in Section V.

Notation: Let $\mathbb{R}$, $\mathbb{R}_+$ and $\mathbb{N}_+$ be the sets of real numbers, non-negative real numbers and non-negative natural numbers, respectively. Denote the transpose of a vector $v$ by $v^\top$ and its norm by $||v||$. Let $||v||_P=\sqrt{v^\top Pv}$ be the weighted norm of the vector $v$ using the matrix $P$. The matrix $P=\operatorname{diag}(P_1,...,P_M)$ denotes a diagonal matrix with the submatrices $P_i, \ i \in \{1,...,M\},$ along its diagonal. Let $\pazocal{X} \times \pazocal{Y}$ denote the cartesian product of the two sets $\pazocal{X}$ and $\pazocal{Y}$. 

\section{PROBLEM FORMULATION}

\label{PF}

We consider a large-scale dynamical system which admit a separable structure and thus, can be decompsed into $M$ subsystems. For each subsystem $i \in \{1,...,M\}$, a set of neighbors is defined comprising subsystem $i$ itself as well as all other subsystems coupled with subsystem $i$ through the dynamics or the constraints. Each subsystem $i$ is described as a discrete-time linear time-invariant system given by
\begin{equation}
\label{sec2_dynamics}
x_i(t+1) = A_{\pazocal{N}_i}x_{\pazocal{N}_i}(t)+B_i u_i(t),
\end{equation}
where $t$ is the time index, $x_i \in \mathbb{R}^{n_i}$, $u_i \in \mathbb{R}^{m_i}$ and $x_{\pazocal{N}_i} \in \mathbb{R}^{n_{N_i}}$ are the state vector of subsystem $i$, the input vector of subsystem $i$ and the state vector of the neighbors of subsystem $i$ respectively. The system matrices
$A_{\pazocal{N}_i} \in \mathbb{R}^{n_i \times n_{N_i}}$ and $B_i \in \mathbb{R}^{n_i \times m_i}$ are assumed to be known.
The state and input constraint sets of each subsystem are given by
\begin{equation}
\label{sec2_constraints}
\begin{aligned}
x_{\pazocal{N}_i}(t) \in \pazocal{X}_{\pazocal{N}_i} &= \{ x_{\pazocal{N}_i} \in \mathbb{R}^{n_{N_i}}: G_{\pazocal{N}_i} x_{\pazocal{N}_i} \leq g_{\pazocal{N}_i} \}, \\
u_i(t) \in \pazocal{U}_i &= \{ u_i \in \mathbb{R}^{m_i}: H_i u_i \leq h_i \}, \\	
\end{aligned}
\end{equation}
where the constraints matrices $G_{\pazocal{N}_i} \in \mathbb{R}^{q_i \times n_{N_i}}$, $H_i \in \mathbb{R}^{r_i \times m_i}$ and vectors $g_{\pazocal{N}_i} \in \mathbb{R}^{q_i}$, $h_i \in \mathbb{R}^{r_i}$ are assumed to be known. We assume that the inputs of the different subsystems are coupled neither through the dynamics, nor through the constraints; indeed this assumption can be imposed without loss of generality, because inputs can always be decoupled by introducing new auxiliary variables \citep{darivianakis2019distributed}.

\begin{assumption}
	The sets $\pazocal{X}_{\pazocal{N}_i}$ and $\pazocal{U}_i$ are convex sets with the origin in their interior.
\end{assumption}

Our main aim is to regulate the system to the origin. We therefore impose a quadratic cost function in the states and the inputs. To maintain the distributed structure of the optimal control problem, the local cost function of subsystem $i$ is assumed to be a function of the states of the neighbors of subsystem $i$ and the inputs of subsystem $i$. Let $T \in \mathbb{N}_+$ be the prediction horizon and define $x_{\pazocal{N}_i}(\pazocal{T})=[x_{\pazocal{N}_i}(0)^\top,..,x_{\pazocal{N}_i}(t)^\top,..,x_{\pazocal{N}_i}(T)^\top]^\top$ and $u_{i}(\pazocal{T})=[u_{i}(0)^\top,..,u_{i}(t)^\top,..,u_{i}(T)^\top]^\top$. Therefore, the local cost function of subsystem $i$ is designed to be
\begin{equation}
\begin{aligned}
\label{sec2_cost}
J_i\left(x_{\pazocal{N}_i}(\pazocal{T}),u_i(\pazocal{T})\right) = &\sum_{t=0}^{T-1} \left[
x_{\pazocal{N}_i}(t)^\top Q_{\pazocal{N}_i} x_{\pazocal{N}_i}(t)
+ u_i(t)^\top R_i u_i(t) \right] \\
&+ x_i(T)^\top P_i x_i(T),
\end{aligned}
\end{equation}
where $Q_{\pazocal{N}_i} \in \mathbb{R}^{n_{N_i} \times n_{N_i}}$ and $R_i \in \mathbb{R}^{m_i \times m_i}$ are the local cost function matrices and $P_i \in \mathbb{R}^{n_i \times n_i}$ is the local terminal cost matrix. The matrices $Q_{\pazocal{N}_i}$, $R_i$ and $P_i$ are assumed to be known.

Denoting the global state and input vectors of the whole system as $x=[x_1^\top,...,x_M^\top]^\top \in \mathbb{R}^n$ and $u=[u_1^\top,...,u_M^\top]^\top \in \mathbb{R}^m$ respectively, the mappings $U_i \in \{0,1\}^{n_i \times n}$, $W_{\pazocal{N}_i} \in \{0,1\}^{ n_{N_i} \times n}$ and $V_i \in \{0,1\}^{m_i \times m}$ can be defined to relate the local variables of subsystem $i$ to the global variables as follows
\begin{equation}
\begin{aligned}
\label{sec2_map}
x_i &= U_i x, \\
x_{\pazocal{N}_i} &= W_{\pazocal{N}_i}x, \\
u_i &= V_i u.
\end{aligned}
\end{equation}

To ensure the asymptotic stability of the closed-loop system and the recursive feasibility of the proposed distributed MPC, the final state $x_i(T)$ of each subsystem $i$ is constrained to lie in an ellipsoidal terminal set as follows
\begin{equation}
\label{sec2_ter}
x_i(T) \in \pazocal{X}_{f,i} = \{x_i \in \mathbb{R}^{n_i} : (x_i-c_i)P_i(x_i-c_i) \leq \alpha_i \},
\end{equation}
where $\alpha_i \in \mathbb{R}$ represents the size of the terminal set and $c_i \in \mathbb{R}^{n_i}$ represents the center of the terminal set. This ellipsoidal terminal set is required to be invariant under the terminal controller $u_{f,i}=K_{\pazocal{N}_i}x_i$. Thus, assuming that $\mathcal{X}_{f,i}(K_{\pazocal{N}_i})$ is the set of ellipsoidal terminal sets which are invariant under the terminal controller $K_{\pazocal{N}_i}$, we impose the constraint
\begin{equation}
\label{extra}
\pazocal{X}_{f,i} \in \mathcal{X}_{f,i}(K_{\pazocal{N}_i}).
\end{equation}
We assume that the terminal controller $K_{\pazocal{N}_i}$ and the matrix $P_i$ have been designed off-line and we seek $c_i$ and $\alpha_i$ online such that $\pazocal{X}_{f,i}$ satisfies \eqref{extra}.


In conclusion, the global cooperative online optimal control problem is formulated as
\begin{equation}
\label{sec2_ocp}
\begin{aligned}
& \min \sum_{i=1}^M J_i(x_{\pazocal{N}_i}(\pazocal{T}),u_i(\pazocal{T})) \\
& s.t. \left\{		
\begin{aligned}
& \left. 
   \begin{aligned} 
   		& x_i(t+1) = A_{\pazocal{N}_i}x_{\pazocal{N}_i}+B_i u_i,\\ 
   		& x_{\pazocal{N}_i}(t) \in \pazocal{X}_{\pazocal{N}_i},\\ 
   		& u_i(t) \in \pazocal{U}_i,\\ 
   \end{aligned}  
\right\} 
\begin{aligned}
\forall t \in \{0,...,T\} \\
\forall i \in \{1,...,M\}
\end{aligned} \\
&
\left.
\begin{aligned}
& x_{\pazocal{N}_i}(0) = x_{\pazocal{N}_i,0}, \\
& x_i(T) \in \pazocal{X}_{f,i}, \\
& \pazocal{X}_{f,i} \in \mathcal{X}_{i,f}(K_{\pazocal{N}_i}),
\end{aligned} 
\quad \quad \quad \ 
\right\} 
\begin{aligned}
\forall i \in \{1,...,M\}
\end{aligned} \\
\end{aligned}
\right.
\end{aligned}
\end{equation}
where $x_{\pazocal{N}_i,0} \in \mathbb{R}^{n_i}$ is the current state of the neighbors of subsystem $i$. The decision variables of this optimal control problem are the predicted state trajectory $x_i(t)$ for all $\ i \in \{1,...,M\} \ \text{and} \ t \in \{1,...,T\}$, the predicted input trajectory $u_i(t)$ for all $ \ i \in \{1,...,M\} \ \text{and} \ t \in \{1,...,T\}$, the terminal set size $\alpha_i$ for all $\ i \in \{1,...,M\}$ and the terminal set center $c_i$ for all $\ i \in \{1,...,M\}$. On the other side, the systems matrices $A_{\pazocal{N}_i}$, $B_i$, the constraint matrices $G_{\pazocal{N}_i}$, $H_i$, the constraint vectors $g_{\pazocal{N}_i}$, $h_i$, the cost function matrices $Q_{\pazocal{N}_i}$, $R_i$, the terminal cost matrix $P_i$ and the terminal controller $K_{\pazocal{N}_i}$ are all known for all $i \in \{1,...,M\}$. The last constraint in \eqref{sec2_ocp} is ensured by means of convex optimization tools in the next section.

\section{Distributed MPC Scheme}

\label{section3}

In the above MPC formulation, the terminal cost matrix and the terminal controller need to be computed appropriately offline to ensure asymptotic stability and recursive feasibility. To compute these terminal ingredients, we follow the method in \cite{conte2012distributed,conte2016distributed}. This method is briefly outlined in Section \ref{OFF} for completeness.

We then modify the online optimal control problem (\ref{sec2_ocp}) by replacing the last constraint with a set of other constraints on the terminal set size and center to ensure positive invariance. Finally, the modified optimal control problem is then relaxed to enlarge the region of attraction of the proposed distributed MPC scheme.

\subsection{Offline Phase}

\label{OFF}

We recall how the terminal cost matrices $P_i$ and the terminal controllers $K_{\pazocal{N}_i}$ for all subsystems $i \in \{1,...,M\}$ can be determined by solving a semidefinite program. This program is mainly based on the idea of defining the terminal cost matrices $P_i$ such that $P=\operatorname{diag}(P_1,..,P_i,..,P_M)  \in \mathbb{R}^{n \times n}$ is a Lyapunov matrix of \eqref{sec2_dynamics} under the terminal controller $u_{f,i}$. With this choice, the terminal controller is stabilizing and the terminal costs upper bounds the infinite horizon cost (\cite{conte2012distributed,conte2016distributed}).
Consider the matrices  
$\Gamma_{\pazocal{N}_i} \in \mathbb{R}^{n_{N_i} \times n_{N_i}}$,  
$S_i \in \mathbb{R}^{n_i \times n_i}$ and 
$S=\operatorname{diag}(S_1,..,S_i,..,S_M) \in \mathbb{R}^{n \times n}$.
Define 
$E=P^{-1}$, 
$E_{\pazocal{N}_i}=W_{\pazocal{N}_i} P^{-1} W_{\pazocal{N}_i}^\top$, 
$E_i=U_{\pazocal{N}_i} P U_{\pazocal{N}_i}^\top$,  
$Y_{\pazocal{N}_i}=K_{\pazocal{N}_i} E_{\pazocal{N}_i}$, 
$H_{\pazocal{N}_i}=E_{\pazocal{N}_i} \Gamma_{\pazocal{N}_i} E_{\pazocal{N}_i}$ and 
$S_{\pazocal{N}_i} = W_{\pazocal{N}_i} S^{-1} W_{\pazocal{N}_i}^\top$.
The semi-definite program is formulated as follows,
\begin{equation}
\label{sec3_offline}
\begin{aligned}
& \max \ \sum_{i=1}^{M} \operatorname{trace} (E_i) \\
& s.t. \left\{		
\begin{aligned}
& E_i \geq 0, \ \forall i \in \{1,...,M\},\\
& \eqref{sec3_suboff} \ \text{holds}, \ \forall i \in \{1,...,M\},\\
& H_{\pazocal{N}_i} \leq S_{\pazocal{N}_i}, \ \forall i \in \{1,...,M\},\\
&\sum_{j \in \pazocal{N}_i} U_i W_{\pazocal{N}_j}^\top S_{\pazocal{N}_j} W_{\pazocal{N}_j} U_i^\top \leq 0, \ \forall i \in \{1,...,M\}. \\
\end{aligned}
\right.
\end{aligned}
\end{equation}
where the desision variables are $E$, $S$, $Y_{\pazocal{N}_i}$ and $H_{\pazocal{N}_i}$ for all $i \in \{1,...,M\}$ and the LMI \eqref{sec3_suboff} is given overleaf in single column. The terminal cost matrices and the terminal controllers can then be computed as $P_i=E_i^{-1}$ and $K_{\pazocal{N}_i}=Y_{\pazocal{N}_i} E_{\pazocal{N}_i}^{-1}$.

\begin{table*}
	\normalsize
	\begin{equation}
	\label{sec3_suboff}
	\begin{bmatrix}
	W_i U_i^\top E_i U_i W_i^\top + H_{\pazocal{N}_i} & E_{\pazocal{N}_i} A_{\pazocal{N}_i}^\top + Y_{\pazocal{N}_i}^\top B_i^\top & E_{\pazocal{N}_i} Q_{\pazocal{N}_i}^\top & Y_{\pazocal{N}_i}^\top R_i^{1/2} \\
	A_{\pazocal{N}_i} E_{\pazocal{N}_i} + B_i Y_{\pazocal{N}_i} & E_i & 0 & 0 \\
	Q_{\pazocal{N}_i}^{1/2} E_{\pazocal{N}_i} & 0 & I & 0 \\
	R_i^{1/2}  Y_{\pazocal{N}_i} & 0 & 0 & I
	\end{bmatrix} \geq 0.
	\end{equation}
\end{table*}

\subsection{Online Phase Modification}

\label{ON_MOD}

Recall that the final state $x_i(T)$ has to satify the constraint
\begin{equation}
\label{sec3_tc}
(x_i(T)-c_i)^\top P_i (x_i(T)-c_i) \leq \alpha_i.
\end{equation}
By means of the Schur complement (\cite{boyd1994linear}), an equivalent form to constraint \eqref{sec3_tc} can be reformulated as
\begin{equation}
\label{sec3_tcMod}
\begin{bmatrix}
P^{-1}_i \alpha^{1/2}_i & x_i(T)-c_i \\
(x_i(T)-c_i)^\top & \alpha^{1/2}_i \\
\end{bmatrix}
\geq 0.
\end{equation}

For the closed-loop system to be asymptotically stable, the local terminal sets $\pazocal{X}_{f,i}, \ i \in \{1,...,M\}$ have to be positively invariant \citep{darivianakis2019distributed}. The following proposition shows the conditions to ensure the positive invariance of the terminal sets.
\begin{prop}
	[\cite{darivianakis2019distributed}]
	\label{prop1}
	Define the sets \newline
	$\pazocal{X}_{\pazocal{N}_i,f}$ $= \times_{j \in \pazocal{N}_i} \pazocal{X}_{j,f} $. Each local terminal set $\pazocal{X}_{f,i}$ is positively invariant if for each $i \in \{1,...,M\}$ and for all $ x_{\pazocal{N}_i} \in \pazocal{X}_{\pazocal{N}_i,f}$,
	\begin{subequations}
		\begin{gather}
		\label{yorgos1}
		(A_{\pazocal{N}_i}+B_i K_{\pazocal{N}_i})x_{\pazocal{N}_i} \in \pazocal{X}_{f,i}, \\
		\label{yorgos2}
		x_{\pazocal{N}_i} \in \pazocal{X}_{\pazocal{N}_i}, \\
		\label{yorgos3}
		K_{\pazocal{N}_i}x_{\pazocal{N}_i} \in \pazocal{U}_{i}.
		\end{gather}
	\end{subequations} 
	Consequently, the global terminal set $\pazocal{X}_{f} = \times_{i \in \{1,...,M\}} \pazocal{X}_{f,i}$ is positively invariant.
\end{prop}
Condition (\ref{yorgos1}) ensures that the terminal set $\pazocal{X}_{f,i}$ is invariant. Whereas, conditions (\ref{yorgos2}) and (\ref{yorgos3}) show that all the state and input constraints are satisfied inside the terminal set respectively. In the sequel, LMIs are derived for each of the conditions in Proposition \ref{prop1}. Embedding these LMIs in the online optimal control problem \eqref{sec2_ocp} guarantees the positive invariance of the terminal set. The derived LMIs depend on the following quantities: $\alpha=\operatorname{diag}(\alpha_1 I_{n_1},..,\alpha_i I_{n_i},..,\alpha_M I_{n_M})$, $c=[c_1^\top,..,c_i^\top,..,c_M^\top]^\top$, $\alpha_{\pazocal{N}_i}=W_{\pazocal{N}_i} \alpha W_{\pazocal{N}_i}^\top$ and $c_{\pazocal{N}_i}=W_{\pazocal{N}_i} c$.

Condition (\ref{yorgos1}) can be represented using an LMI as shown in the following proposition; the inequalities (\ref{moved},\ref{long2},\ref{long3},\ref{longeq}) to which we refer in this proposition are found overleaf in single columns.

\begin{prop}
	\label{prop_inv}
	For each subsystem $i \in \{1,...,M\}$, the terminal set invariance condition
	\begin{equation}
	\begin{aligned}
	\label{mpc_inv1}
	[(A_{\pazocal{N}_i} + &B_i K_{\pazocal{N}_i}) x_{\pazocal{N}_i} - c_i]^\top P_i [(A_{\pazocal{N}_i} +B_i K_{\pazocal{N}_i}) x_{\pazocal{N}_i} - c_i] \leq \alpha_i, \\
	&\forall j \in \pazocal{N}_i, \ x_j \ni (x_j-c_j)^\top P_j (x_j-c_j) \leq \alpha_j,
	\end{aligned}
	\end{equation}
	holds if there exist $\lambda_{ij}\geq0$ such that (\ref{longeq}) holds.
\end{prop}

\begin{pf}
	Define an auxiliary vector $s_i \in \mathbb{R}^{n_i}$ for each subsystem's state vector $x_i$ as follows,
	\begin{equation} 
	x_i = c_i + \alpha_i^{1/2} s_i. \label{aux}
	\end{equation}
	By concatenation, the following relation also holds
	\begin{equation}
	\label{con_aux}
	x_{\pazocal{N}_i} = c_{\pazocal{N}_i} + \alpha_{\pazocal{N}_i}^{1/2} s_{\pazocal{N}_i}.
	\end{equation}
	By substituting these auxiliary vectors in (\ref{mpc_inv1}), the invariance condition is written as
	$$
	\begin{aligned}
	& s_{\pazocal{N}_i}^\top (A_{\pazocal{N}_i} \alpha_{\pazocal{N}_i}^{1/2}+B_i K_{\pazocal{N}_i} \alpha_{\pazocal{N}_i}^{1/2})^\top P_i (A_{\pazocal{N}_i} \alpha_{\pazocal{N}_i}^{1/2}+B_i K_{\pazocal{N}_i} \alpha_{\pazocal{N}_i}^{1/2}) s_{\pazocal{N}_i}
	\\ & +
	[(A_{\pazocal{N}_i} + B_i K_{\pazocal{N}_i}) c_{\pazocal{N}_i} - c_i]^\top P_i (A_{\pazocal{N}_i} \alpha_{\pazocal{N}_i}^{1/2}+B_i K_{\pazocal{N}_i} \alpha_{\pazocal{N}_i}^{1/2}) s_{\pazocal{N}_i}
	\\ & +
	[(A_{\pazocal{N}_i} + B_i K_{\pazocal{N}_i}) c_{\pazocal{N}_i} - c_i]^\top P_i
	[(A_{\pazocal{N}_i} + B_i K_{\pazocal{N}_i}) c_{\pazocal{N}_i} - c_i]
	\leq \alpha_i,
	\end{aligned}
	$$
	$$
	\forall j \in \pazocal{N}_i, \ s_j \ni s_j^\top P_j s_j \leq 1.
	$$
	Using the mapping equations in (\ref{sec2_map}) and multiplying the above equation by $\alpha_i^{-1/2}$ gives the condition \eqref{moved}.
	\begin{table*}
		\normalsize
			\begin{equation}
			\label{moved}
		\begin{aligned}
		&s_{\pazocal{N}_i}^\top (A_{\pazocal{N}_i} \alpha_{\pazocal{N}_i}^{1/2}+B_i K_{\pazocal{N}_i} \alpha_{\pazocal{N}_i}^{1/2})^\top P_i \alpha_i^{-1/2} (A_{\pazocal{N}_i} \alpha_{\pazocal{N}_i}^{1/2}+B_i K_{\pazocal{N}_i} \alpha_{\pazocal{N}_i}^{1/2}) s_{\pazocal{N}_i}
		\\ & \quad \quad +
		(A_{\pazocal{N}_i} + B_i K_{\pazocal{N}_i}) c_{\pazocal{N}_i} - c_i]^\top P_i \alpha_i^{-1/2} (A_{\pazocal{N}_i} \alpha_{\pazocal{N}_i}^{1/2}+B_i K_{\pazocal{N}_i} \alpha_{\pazocal{N}_i}^{1/2}) s_{\pazocal{N}_i}
		\\ & \quad \quad \quad \quad +
		[(A_{\pazocal{N}_i} + B_i K_{\pazocal{N}_i}) c_{\pazocal{N}_i} - c_i]^\top P_i \alpha_i^{-1/2}
		[(A_{\pazocal{N}_i} + B_i K_{\pazocal{N}_i}) c_{\pazocal{N}_i} - c_i]
		\leq \alpha_i^{1/2}
		, \quad
		\forall j \in \pazocal{N}_i, \ s_{\pazocal{N}_i} \ni s_{\pazocal{N}_i}^\top P_{ij} s_{\pazocal{N}_i} \leq 1.
		\end{aligned}
		\end{equation}
			\begin{center}
			$\displaystyle\left\Downarrow\vphantom{\int/4}\right.$
		\end{center}
	\end{table*}
	By applying the S-procedure \citep{boyd1994linear} to \eqref{moved}, the invariance condition for each subsystem $i \in \{1,...,M\}$ holds if there exist $ \lambda_{ij} \geq 0, j \in \pazocal{N}_i$ such that (\ref{long2}) holds. Equation (\ref{long2}) can be rearranged as shown in (\ref{long3}). Applying Schur's complement \citep{boyd1994linear} to (\ref{long3}) leads to the linear matrix inequality (\ref{longeq}). 
\end{pf}

\begin{table*}
	\normalsize
	\begin{equation}
	\label{long2}
	\begin{aligned}
	\sum_{j\in\pazocal{N}_i}\lambda_{ij}
	\begin{bmatrix}
	P_{ij} & 0 \\
	0 & -1
	\end{bmatrix}
	-
	&
	\left[
	\begin{matrix}
	(A_{\pazocal{N}_i} \alpha_{\pazocal{N}_i}^{1/2}+B_i K_{\pazocal{N}_i} \alpha_{\pazocal{N}_i}^{1/2})^\top P_i \alpha_i^{-1/2} (A_{\pazocal{N}_i} \alpha_{\pazocal{N}_i}^{1/2}+B_i K_{\pazocal{N}_i} \alpha_{\pazocal{N}_i}^{1/2}) 
	\\
	(A_{\pazocal{N}_i} + B_i K_{\pazocal{N}_i}) c_{\pazocal{N}_i} - c_i]^\top P_i \alpha_i^{-1/2} (A_{\pazocal{N}_i} \alpha_{\pazocal{N}_i}^{1/2}+B_i K_{\pazocal{N}_i} \alpha_{\pazocal{N}_i}^{1/2})
	\\
	\end{matrix}
	\right.
	\\ &
	\left.
	\begin{matrix}
	(A_{\pazocal{N}_i} \alpha_{\pazocal{N}_i}^{1/2}+B_i K_{\pazocal{N}_i} \alpha_{\pazocal{N}_i}^{1/2})^\top
	P_i \alpha_i^{-1/2}
	(A_{\pazocal{N}_i} + B_i K_{\pazocal{N}_i}) c_{\pazocal{N}_i} - c_i]
	\\
	[(A_{\pazocal{N}_i} + B_i K_{\pazocal{N}_i}) c_{\pazocal{N}_i} - c_i]^\top P_i \alpha_i^{-1/2}
	[(A_{\pazocal{N}_i} + B_i K_{\pazocal{N}_i}) c_{\pazocal{N}_i} - c_i] - \alpha_i^{1/2}
	\end{matrix}
	\right] \geq 0
	\end{aligned}
	\end{equation}
	\begin{center}
		$\displaystyle\left\Downarrow\vphantom{\int}\right.$
	\end{center}
\end{table*}

\begin{table*}
	\normalsize
	\begin{equation}
	\label{long3}
	\begin{aligned}
	&
	\begin{bmatrix}
	\sum_{j \in \pazocal{N}_i} \lambda_{ij} P_{ij} & 0 \\
	0 & \alpha_i^{1/2} - \sum_{j \in \pazocal{N}_i} \lambda_{ij}
	\end{bmatrix}
	-
	\begin{bmatrix}
	(A_{\pazocal{N}_i} \alpha_{\pazocal{N}_i}^{1/2}+B_i K_{\pazocal{N}_i} \alpha_{\pazocal{N}_i}^{1/2})^\top \\
	(A_{\pazocal{N}_i} + B_i K_{\pazocal{N}_i}) c_{\pazocal{N}_i} - c_i]^\top
	\end{bmatrix}
	P_i \alpha_i^{-1/2}
	\begin{bmatrix}
	(A_{\pazocal{N}_i} \alpha_{\pazocal{N}_i}^{1/2}+B_i K_{\pazocal{N}_i} \alpha_{\pazocal{N}_i}^{1/2})
	&
	[(A_{\pazocal{N}_i} + B_i K_{\pazocal{N}_i}) c_{\pazocal{N}_i} - c_i]
	\end{bmatrix}			
	\geq 0
	\end{aligned}
	\end{equation}
	\begin{center}
		$\displaystyle\left\Downarrow\vphantom{\int}\right.$
	\end{center}
\end{table*}

\begin{table*}
	\normalsize
	\begin{equation}
	\label{longeq}
	\begin{bmatrix}
	P_i^{-1} \alpha_i^{1/2} & (A_{\pazocal{N}_i} \alpha_{\pazocal{N}_i}^{1/2}+B_i K_{\pazocal{N}_i} \alpha_{\pazocal{N}_i}^{1/2})
	&
	[(A_{\pazocal{N}_i} + B_i K_{\pazocal{N}_i}) c_{\pazocal{N}_i} - c_i]
	\\
	(A_{\pazocal{N}_i} \alpha_{\pazocal{N}_i}^{1/2}+B_i K_{\pazocal{N}_i} \alpha_{\pazocal{N}_i}^{1/2})^\top & \sum_{j \in \pazocal{N}_i} \lambda_{ij} P_{ij} & 0 \\
	(A_{\pazocal{N}_i} + B_i K_{\pazocal{N}_i}) c_{\pazocal{N}_i} - c_i]^\top & 0 & \alpha_i^{1/2} - \sum_{j \in \pazocal{N}_i} \lambda_{ij}
	\end{bmatrix}
	\geq 0
	\end{equation}
\end{table*}

\ \\

Condition (\ref{yorgos2}) can be represented as an LMI as shown in the following proposition.

\begin{prop}
	\label{prop_sc1}
	Denote the $k^{\mbox{\em th}}$ row of the matrix $G_{\pazocal{N}_i}$ by $G^k_{\pazocal{N}_i}$ and the $k^{\mbox{\em th}}$ element of the vector $g_{\pazocal{N}_i}$ by $g^k_{\pazocal{N}_i}$. For each subsystem $i \in \{1,...,M\}$, the state constraint $k \in \{1,2,...,q_i\}$
	\begin{equation}
	\label{mpc_sc1} 
	G_{\pazocal{N}_i}^k x_{\pazocal{N}_i} \leq g_{\pazocal{N}_i}^k, \quad
	\forall j \in \pazocal{N}_i, \ x_j\ni(x_j-c_j)^\top P_j (x_j-c_j) \leq \alpha_j,
	\end{equation}
	holds if  there exist $\tau_{ij}^k \geq 0$ such that
	\begin{equation}
	\label{mpc_sc2}
	\begin{bmatrix}
	g_{\pazocal{N}_i}^k & G_{\pazocal{N}_i}^{k}
	\alpha_{\pazocal{N}_i}^{1/2}
	&
	G_{\pazocal{N}_i}^{k}
	c_{\pazocal{N}_i}
	\\
	\alpha_{\pazocal{N}_i}^{1/2}
	G_{\pazocal{N}_i}^{k^\top} & \sum_{j \in \pazocal{N}_i} \tau_{ij}^k P_{ij} & 0 \\
	c_{\pazocal{N}_i}^\top G_{\pazocal{N}_i}^{k^\top} & 0 & g_{\pazocal{N}_i}^k - \sum_{j \in \pazocal{N}_i} \tau_{ij}^k
	\end{bmatrix}
	\geq 0.
	\end{equation}
\end{prop}

\begin{pf}
	Consider the auxiliary vectors $s_i$ and the concatenated auxiliary vectors $s_{\pazocal{N}_i}$ defined in \eqref{aux}, \eqref{con_aux}. Substituting these auxiliary vectors in (\ref{mpc_sc1}), the state constraints become
	$$
	G_{\pazocal{N}_i}^k (c_{\pazocal{N}_i}+\alpha_{\pazocal{N}_i}^{1/2} s_{\pazocal{N}_i}) \leq g_{\pazocal{N}_i}^k, \quad
	\forall  j \in \pazocal{N}_i, \ s_j \ni s_j^\top P_j s_j \leq 1.
	$$
	A more conservative approximation of the above implication can be obtained by squaring the above inequality following \cite{darivianakis2019distributed}. Using the mapping equations in (\ref{sec2_map}), the resulting implication is given by
	$$
	(c_{\pazocal{N}_i}+\alpha_{\pazocal{N}_i}^{1/2} s_{\pazocal{N}_i})^\top 
	G_{\pazocal{N}_i}^{k^\top}
	g_{\pazocal{N}_i}^{k^{-1}}
	G_{\pazocal{N}_i}^k (c_{\pazocal{N}_i}+\alpha_{\pazocal{N}_i}^{1/2} s_{\pazocal{N}_i}) \leq {g_{\pazocal{N}_i}^k}
	$$
	$$
	\forall j \in \pazocal{N}_i, \ s_{\pazocal{N}_i} \ni s_{\pazocal{N}_i}^\top P_{ij} s_{\pazocal{N}_i} \leq 1.
	$$
	By applying the S-procedure \citep{boyd1994linear} to the above implication, the state constraints of each subsystem $i \in \{1,...,M\}$ are satisfied inside the terminal set if there exists $\tau_{ij}^k \geq 0$ such that the following LMI holds.
	$$
	\begin{aligned}
	& \sum_{j\in\pazocal{N}_i}\tau_{ij}^k
	\begin{bmatrix}
	P_{ij} & 0 \\
	0 & -1
	\end{bmatrix}
	-
	\\ &
	\left[
	\begin{matrix}
	\alpha_{\pazocal{N}_i}^{1/2}
	G_{\pazocal{N}_i}^{l^\top}
	g_{\pazocal{N}_i}^{l^{-1}}
	G_{\pazocal{N}_i}^l
	\alpha_{\pazocal{N}_i}^{1/2}
	&
	\alpha_{\pazocal{N}_i}^{1/2}
	G_{\pazocal{N}_i}^{l^\top}
	g_{\pazocal{N}_i}^{l^{-1}}
	G_{\pazocal{N}_i}^l
	c_{\pazocal{N}_i}
	\\
	c_{\pazocal{N}_i}^\top
	G_{\pazocal{N}_i}^{l^\top}
	g_{\pazocal{N}_i}^{l^{-1}}
	G_{\pazocal{N}_i}^l
	\alpha_{\pazocal{N}_i}^{1/2}
	&
	c_{\pazocal{N}_i}^\top
	G_{\pazocal{N}_i}^{l^\top}
	g_{\pazocal{N}_i}^{l^{-1}}
	G_{\pazocal{N}_i}^l
	c_{\pazocal{N}_i}
	-
	{g_{\pazocal{N}_i}^l}
	\end{matrix}
	\right] \geq 0.
	\end{aligned}
	$$
	This LMI can be rearranged and expressed as
	$$
	\begin{aligned}
	& \begin{bmatrix}
	\sum_{j\in\pazocal{N}_i}\tau_{ij}^k P_{ij} & 0 \\
	0 & {g_{\pazocal{N}_i}^l}-\sum_{j\in\pazocal{N}_i}\tau_{ij}
	\end{bmatrix}
	-
	\\ &
	\left[
	\begin{matrix}
	\alpha_{\pazocal{N}_i}^{1/2}
	G_{\pazocal{N}_i}^{l^\top}
	\\
	c_{\pazocal{N}_i}^\top
	G_{\pazocal{N}_i}^{l^\top}
	\end{matrix}
	\right]
	g_{\pazocal{N}_i}^{l^{-1}}
	\left[
	\begin{matrix}
	G_{\pazocal{N}_i}^{l}
	\alpha_{\pazocal{N}_i}^{1/2}
	&
	G_{\pazocal{N}_i}^{l}
	c_{\pazocal{N}_i}
	\end{matrix}
	\right]
	\geq 0.
	\end{aligned}
	$$
	By applying Schur's complement \citep{boyd1994linear} to the above inequality, the LMI in (\ref{mpc_sc2}) is reached.
\end{pf}

\ \\

Condition (\ref{yorgos3}) can be represented using an LMI as shown in the following proposition.

\begin{prop}
	\label{prop_ic1}
	Denote the $l^{\mbox{\em th}}$ row of the matrix $H_{\pazocal{N}_i}$ by $H^l_{\pazocal{N}_i}$ and the $l^{\mbox{\em th}}$ element of the vector $h_{\pazocal{N}_i}$ by $h^l_{\pazocal{N}_i}$. For each subsystem $i \in \{1,...,M\}$ whose neighbors are subsystems $j \in \pazocal{N}_i$, the input constraint $l \in \{1,2,...,r_i\}$
	\begin{equation*}
	H_{\pazocal{N}_i}^l K_{\pazocal{N}_i} x_{\pazocal{N}_i} \leq h_{\pazocal{N}_i}^l, \quad
	\forall j \in \pazocal{N}_i, \ x_j \in (x_j-c_j)^\top P_j (x_j-c_j) \leq \alpha_j,	
	\end{equation*}
	holds if  there exist $\rho_{ij}^l \geq 0$ such that
	\begin{equation}
	\label{sec21_ic}
	\begin{bmatrix}
	h_{\pazocal{N}_i}^l & 			H_{\pazocal{N}_i}^{l} K_{\pazocal{N}_i}
	\alpha_{\pazocal{N}_i}^{1/2}
	&
	H_{\pazocal{N}_i}^{l} K_{\pazocal{N}_i}
	c_{\pazocal{N}_i}
	\\
	\alpha_{\pazocal{N}_i}^{1/2} K_{\pazocal{N}_i}^\top
	H_{\pazocal{N}_i}^{l^\top} & \sum_{j \in \pazocal{N}_i} \rho_{ij}^k P_{ij} & 0 \\
	c_{\pazocal{N}_i}^\top K_{\pazocal{N}_i}^\top H_{\pazocal{N}_i}^{l^\top} & 0 & h_{\pazocal{N}_i}^l - \sum_{j \in \pazocal{N}_i} \rho_{ij}^k
	\end{bmatrix}
	\geq 0.
	\end{equation}
\end{prop}

\begin{pf}
	The proof of this proposition follows the proof of Proposition \ref{prop_sc1} by replacing $\tau_{ij}^k$, $g_{\pazocal{N}_i}^k$ and $G_{\pazocal{N}_i}^k$ with $\rho_{ij}^l$, $h_{\pazocal{N}_i}^l$ and $H_{\pazocal{N}_i}^l K_{\pazocal{N}_i}$ respectively.
\end{pf}

The LMIs derived in \cite{darivianakis2019distributed} are the same as the LMIs (\ref{longeq},\ref{mpc_sc2},\ref{sec21_ic}) evaluated at $c_i=0 \ \forall i \in \{1,2,...,M\}$. Therefore, Propositions \ref{prop_inv}, \ref{prop_sc1} and \ref{prop_ic1} are generalizations of those found in \cite{darivianakis2019distributed}. 
Notice that the center $c_i$ and the size $\alpha_i$ of each local terminal set are considered as decision variables in this setting without affecting the convexity of the problem. However, it is not possible to achieve convex conditions, and thus a convex optimization problem, when considering the terminal control gain $K_{\pazocal{N}_i}$ as a decision variable. This fact is due to the existence of the terms $K_{\pazocal{N}_i} \alpha_{\pazocal{N}_i}^{1/2}$ and $K_{\pazocal{N}_i} c_{\pazocal{N}_i}$ which would result in a nonconvex problem if the gain $K_{\pazocal{N}_i}$ is assumed to be a decision variable.

In conclusion, the online optimal control problem of this distributed MPC with aymmetric adaptive terminal sets is given by
\begin{equation}
\label{mpc_aats}
\begin{aligned}
& \min \ \sum_{i=1}^{M} J_i(x_{N_i}(\pazocal{T}),u_i(\pazocal{T})) \\
& s.t. \left\{		
\begin{aligned}
& x_i(0) = x_{i,0} \ \forall i \in \{1,2,...,M\},\\
& \eqref{sec2_dynamics}, \eqref{sec2_constraints} \ \forall t \in \{0,1,...,T\}, \ \forall i \in \{1,2,...,M\},\\
& \eqref{sec3_tcMod}, \eqref{longeq}  \ \forall i \in \{1,2,...,M\}, \\
& (\ref{mpc_sc2}) \ \forall k \in \{1,2,...,q_i\}, \ \forall i \in \{1,2,...,M\},\\
& (\ref{sec21_ic})  \ \forall l \in \{1,2,...,r_i\}, \ \forall i \in \{1,2,...,M\}.
\end{aligned}
\right.
\end{aligned}
\end{equation}

The following theorem shows that this MPC scheme is recursively feasible and the closed-loop system is asymptotically stable whenever the optimization problem is initially feasible.

\begin{thm}
	The distributed MPC problem with asymmetric adaptive terminal sets is recursively feasible and the closed-loop system under this MPC controller is asymptotically stable.
\end{thm}

\begin{pf}
	The proof of this theorem follows the proof of Theorem 3 in \cite{darivianakis2019distributed}.
\end{pf}

\subsection{Online Phase Relaxation}

The LMIs corresponding to the state and input constraints are derived in Propositions \ref{prop_sc1} and \ref{prop_ic1} by transforming the linear constraints to quadratic ones as in \cite{darivianakis2019distributed}. Using the linear constraints without transforming them into quadratic ones is less conservative. In this section, alternative LMIs are derived in Propositions \ref{prop_sc2} and \ref{prop_ic2} based on the linear constraints. It is found that using the linear constraints, a convex online optimal control problem can still be reached.

Condition (\ref{yorgos2}) can be represented as an LMI (\ref{mpc_sc4}), as shown in the following proposition.

\begin{prop}
	\label{prop_sc2}
	Denote the $k^{\mbox{\em th}}$ row of the matrix $G_{\pazocal{N}_i}$ by $G^k_{\pazocal{N}_i}$ and the $k^{\mbox{\em th}}$ element of the vector $g_{\pazocal{N}_i}$ by $g^k_{\pazocal{N}_i}$. For each subsystem $i \in \{1,...,M\}$, the state constraint $k \in \{1,2,...,q_i\}$
	\begin{equation}
	\label{mpc_sc3}
	G_{\pazocal{N}_i}^k x_{\pazocal{N}_i} \leq g_{\pazocal{N}_i}^k, \quad 
	\forall j \in \pazocal{N}_i, \ x_j \ni (x_j-c_j)^\top P_j (x_j-c_j) \leq \alpha_j, 
	\end{equation}
	holds if  there exist $\sigma_{ij}^k \geq 0$ such that
	\begin{equation}
	\label{mpc_sc4}
	\begin{aligned}
	\begin{bmatrix}
	\sum_{j\in\pazocal{N}_i}\sigma_{ij}^k P_{ij} & \frac{1}{2} \alpha_{\pazocal{N}_i}^{1/2} G_{\pazocal{N}_i}^{k^\top} \\
	\frac{1}{2} G_{\pazocal{N}_i}^k \alpha_{\pazocal{N}_i}^{1/2} & g_{\pazocal{N}_i}^k - 			G_{\pazocal{N}_i}^k   c_{\pazocal{N}_i} - \sum_{j\in\pazocal{N}_i}\sigma_{ij}^k
	\end{bmatrix}
	\geq 0.
	\end{aligned}
	\end{equation}
\end{prop}

\begin{pf}
	Consider the auxiliary vectors $s_i$ and the concatenated auxiliary vectors $s_{\pazocal{N}_i}$ defined in (\ref{aux},\ref{con_aux}). Substituting these auxiliary vectors in (\ref{mpc_sc1}), the state constraints become
	$$
	G_{\pazocal{N}_i}^k (c_{\pazocal{N}_i}+\alpha_{\pazocal{N}_i}^{1/2} s_{\pazocal{N}_i}) \leq g_{\pazocal{N}_i}^k, \quad
	\forall j \in \pazocal{N}_i, \ s_j \ni s_j^\top P_j s_j \leq 1.
	$$
	Using the mapping equations in (\ref{sec2_map}), the above implication can be expressed as
	$$
	G_{\pazocal{N}_i}^k \alpha_{\pazocal{N}_i}^{1/2} s_{\pazocal{N}_i}
	+
	G_{\pazocal{N}_i}^k c_{\pazocal{N}_i}
	\leq h_{\pazocal{N}_i}^l, \quad
	\forall j \in \pazocal{N}_i, \ 
	s_{\pazocal{N}_i}^\top P_{ij} s_{\pazocal{N}_i} \leq 1.
	$$
	Applying the S-procedure \cite{boyd1994linear} to the above implication yields
	$$
	\begin{aligned}
	\sum_{j\in\pazocal{N}_i}\sigma_{ij}^k
	\begin{bmatrix}
	P_{ij} & 0 \\
	0 & -1
	\end{bmatrix}
	-
	&
	\left[
	\begin{matrix}
	0 &
	\frac{1}{2} \alpha_{\pazocal{N}_i}^{1/2}
	G_{\pazocal{N}_i}^{k^\top} 
	\\
	\frac{1}{2} G_{\pazocal{N}_i}^k  \alpha_{\pazocal{N}_i}^{1/2}
	&
	G_{\pazocal{N}_i}^k   c_{\pazocal{N}_i} - g_{\pazocal{N}_i}^k
	\end{matrix}
	\right] \geq 0.
	\end{aligned}
	$$
	Rearranging the above LMI results in (\ref{mpc_sc4}).
\end{pf}

\ \\

Condition (\ref{yorgos3}) can be represented as an LMI (\ref{sec22_ic}), as shown in the following proposition.

\begin{prop}
	\label{prop_ic2}
	Denote the $l^{\mbox{\em th}}$ row of the matrix $H_{\pazocal{N}_i}$ by $H^l_{\pazocal{N}_i}$ and the $l^{\mbox{\em th}}$ element of the vector $h_{\pazocal{N}_i}$ by $h^l_{\pazocal{N}_i}$. For each subsystem $i \in \{1,...,M\}$, the input constraint $l \in \{1,2,...,r_i\}$
	\begin{equation*}
	H_{\pazocal{N}_i}^l K_{\pazocal{N}_i} x_{\pazocal{N}_i} \leq h_{\pazocal{N}_i}^l, \ 
	\forall j \in \pazocal{N}_i, \ x_j \ni (x_j-c_j)^\top P_j (x_j-c_j) \leq \alpha_j,
	\end{equation*}
	holds if  there exist $ \beta_{ij}^l \geq 0$ such that
	\begin{equation}
	\label{sec22_ic}
	\begin{aligned}
	\begin{bmatrix}
	\sum_{j\in\pazocal{N}_i}\beta_{ij}^l P_{ij} & \frac{1}{2} \alpha_{\pazocal{N}_i}^{1/2}
	K_{\pazocal{N}_i}^\top  H_{\pazocal{N}_i}^{l^\top} \\
	\frac{1}{2} H_{\pazocal{N}_i}^l K_{\pazocal{N}_i} \alpha_{\pazocal{N}_i}^{1/2} & h_{\pazocal{N}_i}^l -			H_{\pazocal{N}_i}^l K_{\pazocal{N}_i} c_{\pazocal{N}_i} - \sum_{j\in\pazocal{N}_i}\beta_{ij}^l
	\end{bmatrix}
	\geq 0.
	\end{aligned}
	\end{equation}
\end{prop}

\begin{pf}
	The proof of this proposition follows the proof of Proposition \ref{prop_sc2} by replacing $\sigma_{ij}^k$, $g_{\pazocal{N}_i}^k$ and $G_{\pazocal{N}_i}^k$ with $\beta_{ij}^l$, $h_{\pazocal{N}_i}^l$ and $H_{\pazocal{N}_i}^l K_{\pazocal{N}_i}$ respectively.
\end{pf}

When the LMIs are derived based on the quadratic state and input constraints in Section \ref{ON_MOD}, the variables $c_i$ lie in the off-diagonal terms. When the LMIs are relaxed by using the linear state and input constraints in this section, the variables $c_i$ appear in the diagonal terms. In both cases, the values of the variables $c_i$ are constrained to increase/decrease in some directions. In the former case, the values of the variables $c_i$ are constrained to move along the direction perpendicular to the hyperplane defining the state/input constraint. In the latter case, the values of the variables $c_i$ are constrained to move in the direction perpendicular and pointing towards the hyperplane defining the state/input constraint. Thus, by intuition, the relaxed optimal control problem may indeed have a larger feasible region.  Notice that the invariance LMI (\ref{longeq}) remains the same with the variables $c_i$ appearing in the off-diagonal terms since the invariance condition (\ref{mpc_inv1}) is quadratic by definition. It is worth mentioning that the feasible regions of the proposed scheme and its relaxed version are difficult to compare formally due to the conservativeness introduced by the S procedure. Intuition and simulation results suggest, however, that the feasible region of the relaxed formulation may indeed be larger.

In conclusion, the online optimal control problem of this distrubted MPC with relaxed aymmetric adaptive terminal sets is given by
\begin{equation}
\label{mpc_raats}
\begin{aligned}
& \min \ \sum_{i=1}^{M} J_i(x_{N_i}(\pazocal{T}),u_i(\pazocal{T})) \\
& s.t. \left\{		
\begin{aligned}
& x_{N_i}(0) = x_{N_i,0} \ \forall i \in \{1,2,...,M\}\\
& \eqref{sec2_dynamics}, \eqref{sec2_constraints}, \eqref{sec3_tcMod}, \eqref{longeq}  \ \forall i \in \{1,2,...,M\}\\
& (\ref{mpc_sc4}) \ \forall k \in \{1,2,...,q_i\}, \ \forall i \in \{1,2,...,M\}\\
& (\ref{sec22_ic})  \ \forall l \in \{1,2,...,r_i\}, \ \forall i \in \{1,2,...,M\}
\end{aligned}
\right.
\end{aligned}
,
\end{equation}

The following theorem shows that this MPC scheme is recursively feasible and the closed loop system is asymptotically stable whenever the optimization problem is initially feasible.

\begin{thm}
	The distributed MPC problem with relaxed asymmetric adaptive terminal sets is recursively feasible and the closed-loop system under this MPC controller is asymptotically stable.
\end{thm}

\begin{pf}
	The proof of this theorem follows the proof of Proposition 2 in \cite{darivianakis2019distributed}.
\end{pf}

\subsection{Distributed Implementation}

Although the global cooperative online optimal control problem (\ref{mpc_raats}) is expressed centrally, it is still possible to be solved in a distributed fashion using a distributed optimization technique such as the alternating direction method of multipliers (ADMM) (check \cite{boyd2011distributed} for more details). In this case, the local controller of subsystem $i$ sets initial values for $x_i$, $\alpha_i$ and $c_i$ which are communicated to its neighbors. The local controller then solves a local optimization problem whose optimal solution is $(x_{N_i}^*(\pazocal{T}),u_i^*(\pazocal{T}),\alpha_{N_i}^*,c_{N_i}^*)$ taking into consideration the initial values sent by its neighbors. The controller then communicates $(x_{N_i}^*(\pazocal{T}),\alpha_{N_i}^*, c_{N_i}^*)$ with its neighbors. Finally, the initial values of $x_i$, $\alpha_i$ and $c_i$ are updated based on the communicated optimal solution and sent back to the neighbors. This procedure is repeated until a consensus is reached on the communicated variables. The same procedure holds for the optimal control problem (\ref{mpc_aats}) as well.


\section{SIMULATION RESULTS}

In this section, the effectiveness of the proposed distributed MPC with asymmetric adaptive terminal set \eqref{mpc_aats} (denoted as D-ASYM) and its relaxed version \eqref{mpc_raats} (denoted as D-RLXD) is illustrated by means of a simulation example. These two schemes are compared to the distributed MPC with adaptive terminal set (denoted as D-ADAP) developed in \cite{darivianakis2019distributed} to emphasize their efficacy.

We consider the unstable discrete-time linear time-invariant system
\begin{equation*}
\begin{bmatrix}
{x}_1^+ \\ {x}_2^+ \\
\end{bmatrix}
=
\begin{bmatrix}
2 & 0.5 \\ 0.5 & 2
\end{bmatrix}
\begin{bmatrix}
x_1 \\ x_2
\end{bmatrix}
+
\begin{bmatrix}
1 & 0 \\ 0 & 1
\end{bmatrix}
\begin{bmatrix}
u_1 \\ u_2
\end{bmatrix}.
\end{equation*}
The state and input constraints of this system are represented as
\begin{gather*}
-5 \leq x_i \leq 5, \quad i \in \{1,2\} \\
-0.25 \leq u_i \leq 1, \quad i \in \{1,2\}
\end{gather*}
This system can be decomposed into two subsystems, each of which is coupled with the other through the dynamics. The system and constraint matrices in \eqref{sec2_dynamics}, \eqref{sec2_constraints} can be derived accordingly. The cost function matrices are selected to be $Q_{\pazocal{N}_1}=Q_{\pazocal{N}_2}=0.5I_2$ and $R_1=R_2=0.1$. The terminal cost and controller can then be computed according to (\ref{sec3_offline}).

\begin{figure*}
	\label{res_fig1}
	\includegraphics[scale=0.33]{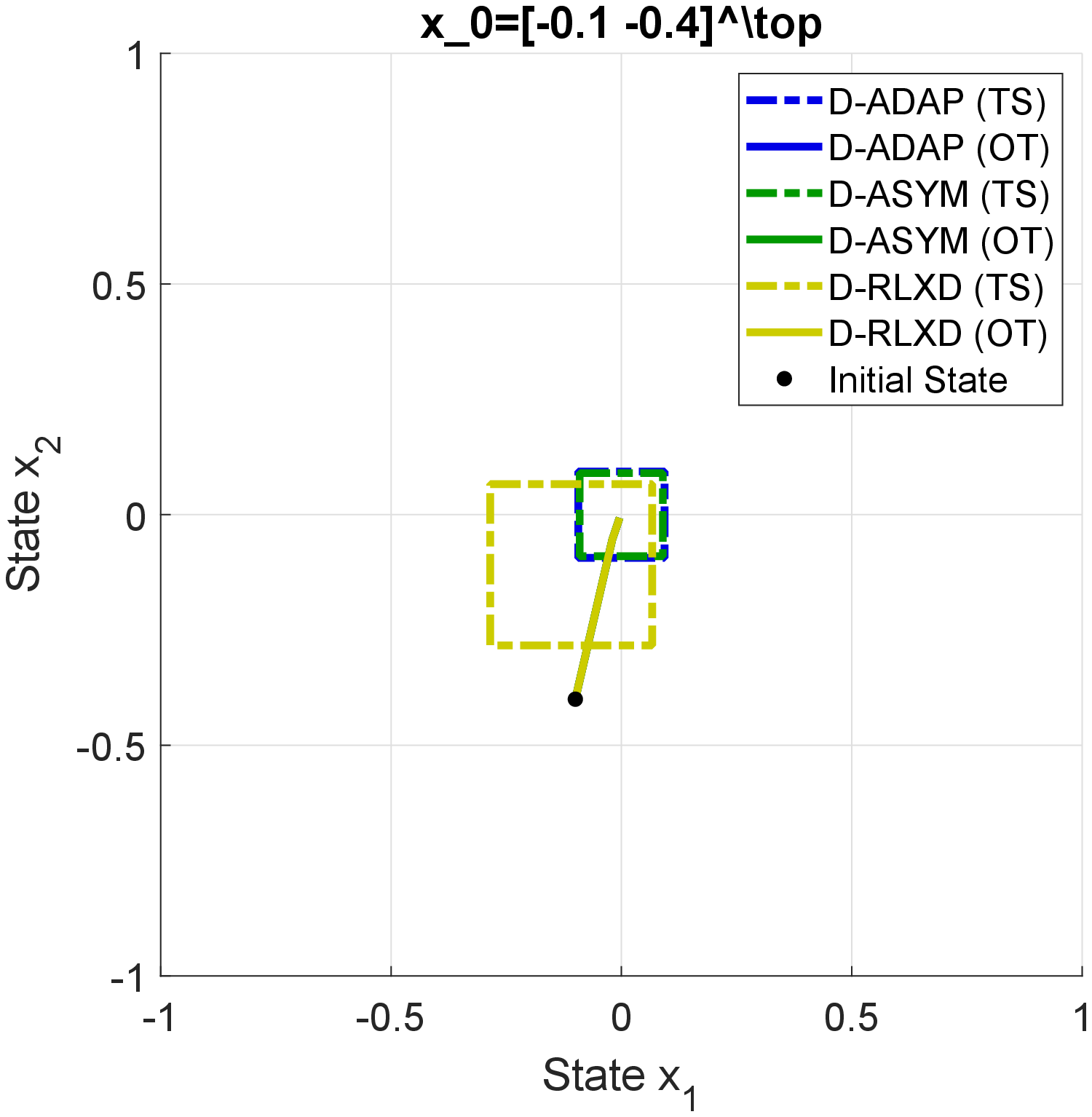}
	\includegraphics[scale=0.33]{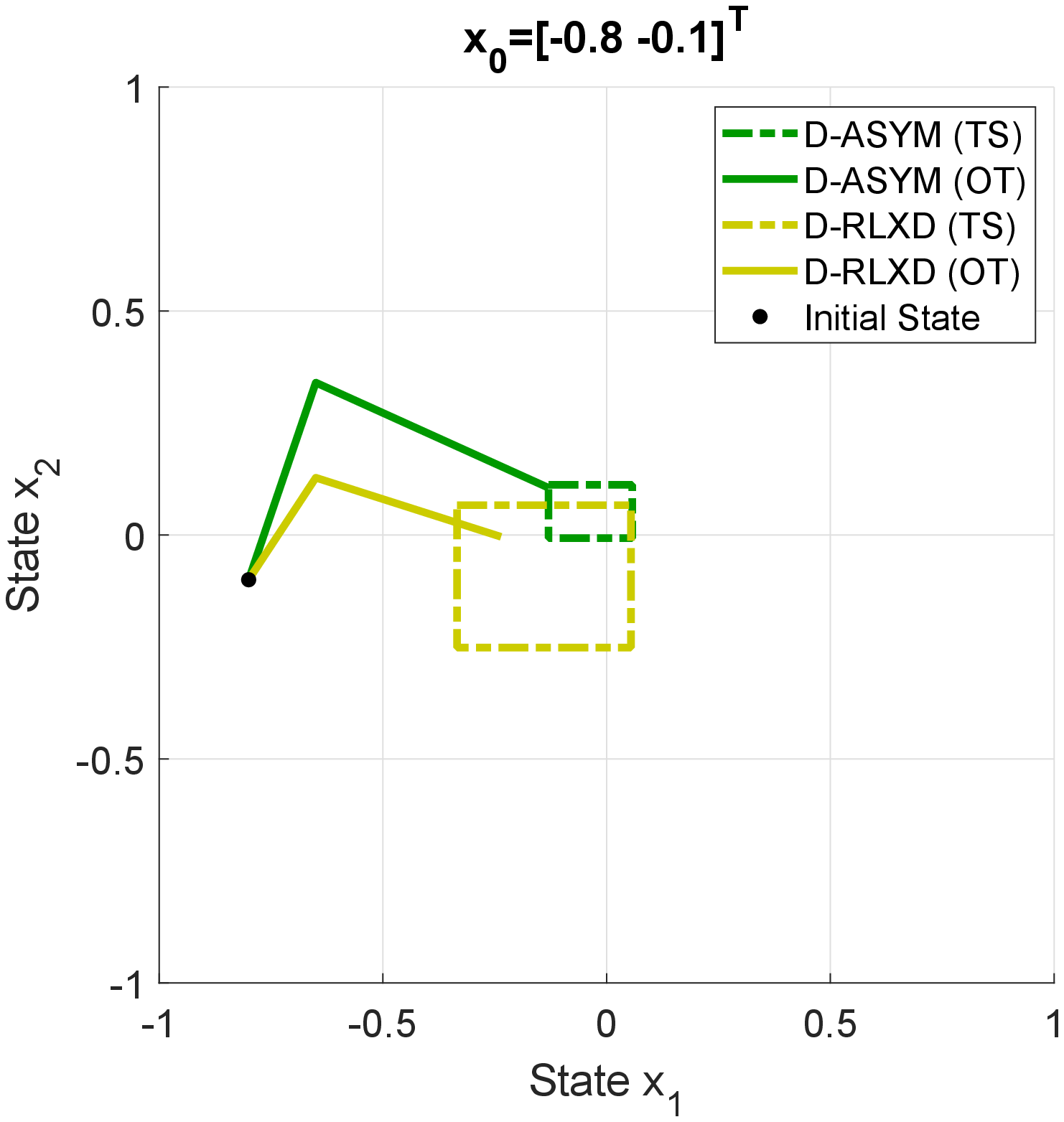}
	\includegraphics[scale=0.33]{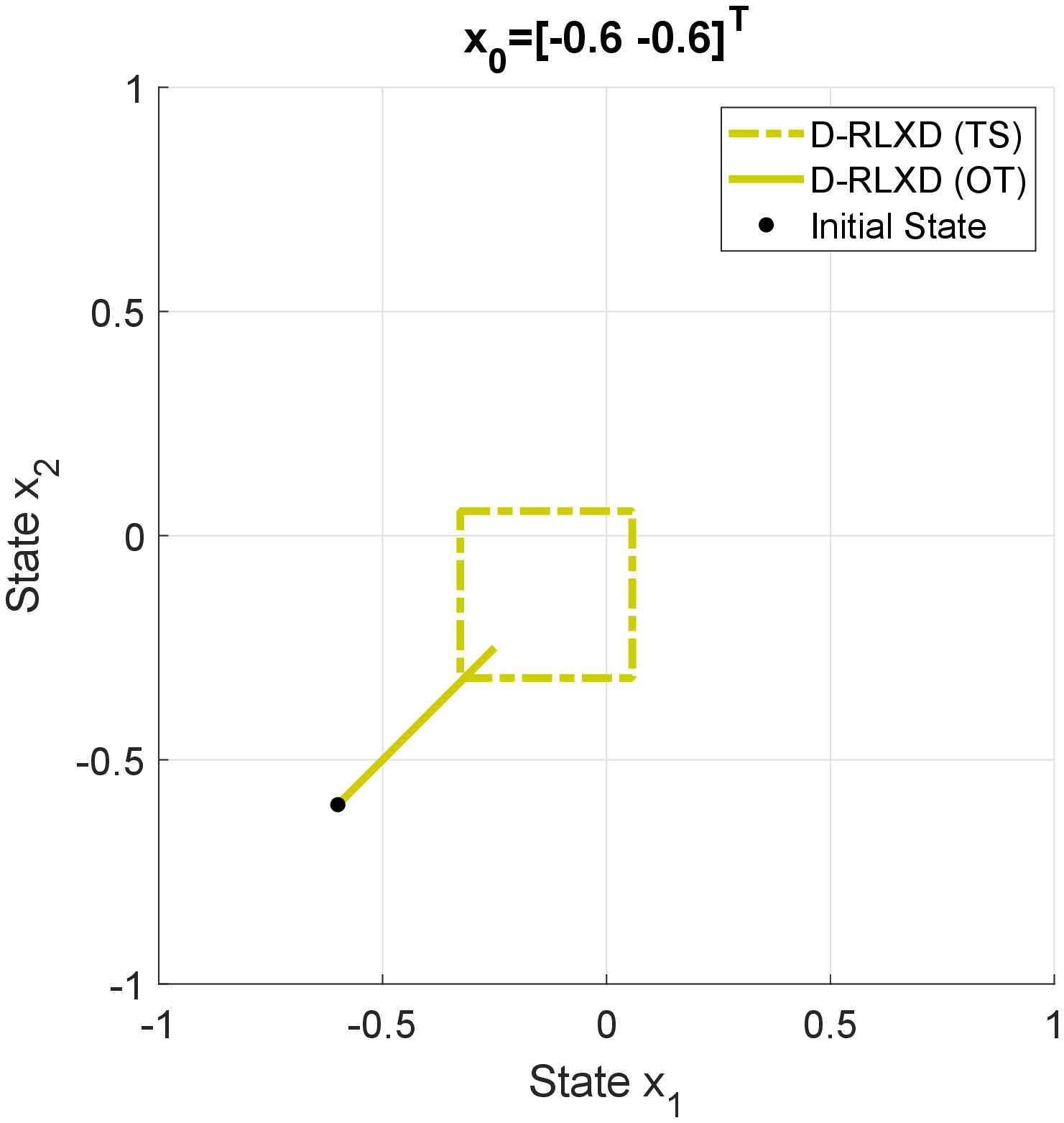}
	\caption{Predicted optimal state trajectories (OT) and terminal sets (TS) of three distributed MPC schemes; D-ADAP (Blue), D-ASYM (Green) and D-RLXD (Yellow), for three different initial conditions and a prediction horizon of $T=2$ when solving the optimization problem once}
\end{figure*}

Figure 1 shows the predicted state trajectory (refered to as OT) and the terminal set (refered to as TS) of the three distributed MPC schemes for three different initial conditions when the optimization problem is solved once. The online optimal control problem is initially feasible for all the schemes when the initial condition is $x_0=[-0.1 \ -0.4]^\top$ and the state trajectories of all the schemes are the same. However, the terminal set of D-RLXD is clearly not centered at the origin and is found to be larger than the terminal sets of the other two schemes. This is because the terminal set is not constrained to be centered at the origin as in D-ADAP and the LMIs derived for D-RLXD are relaxed compared to those for D-ASYM. Notice that the D-ADAP terminal set is partially hidden behind that of D-ASYM in Figure \ref{res_fig1}. Although the terminal set of D-ASYM is almost centered at the origin, this is not necessarily the case as shown when the initial condition is $x_0=[-0.8 \ -0.1]^\top$. In this case, D-ADAP is not initially feasible due to the constraint that the center of its terminal set should be the origin. On the other hand, D-ASYM is initially feasible with the center of its terminal set not located at the origin. Similarly, D-RLXD is also initially feasible and its terminal set is larger than that of D-ASYM. Finally, for the initial condition $x_0=[-0.6 \ -0.6]^\top$, D-RLXD is the only initially-feasible scheme showing that its domain of attraction comprises some parts in the state space that are not included in the domain of attraction of D-ASYM. It is worth mentioning that the terminal set of one scheme is not the same for all the initial conditions and is going to change in the next time steps because the terminal set is determined and updated online. Notice also that the terminal set is described by a rectangle and not an ellipsoid because it is the product of two ellipsoidal sets in one dimension.

Table 1 shows the value of the cost function for the different schemes and initial conditions. When the initial condition is $x_0=[-0.1 \ -0.4]^\top$, the cost of all schemes is the same because the state and input trajectories are the same independently of the scheme applied. In the case of $x_0=[-0.8 \ -0.1]^\top$, the cost of D-ASYM is higher than that of D-RLXD because D-ASYM results in a relatively small terminal set leading to a suboptimal solution. Finally, for $x_0=[-0.6 \ -0.6]^\top$, the cost of D-RLXD is 1.8185.

Figure 2 shows the state and input trajectories for D-ASYM (with the initial condition $x_0=[-0.8 \ -0.1]^\top$) and D-RLXD (with the initial condition $x_0=[-0.6 \ -0.6]^\top$, when the other two schemes are already infeasible) when the optimization problem is solved recursively. Figure 2 emphasizes the fact that the two schemes D-ASYM and D-RLXD are recursively feasible and their closed loop system is asymptotically stable.

\begin{figure*}
	\centering
	\includegraphics[scale=0.35]{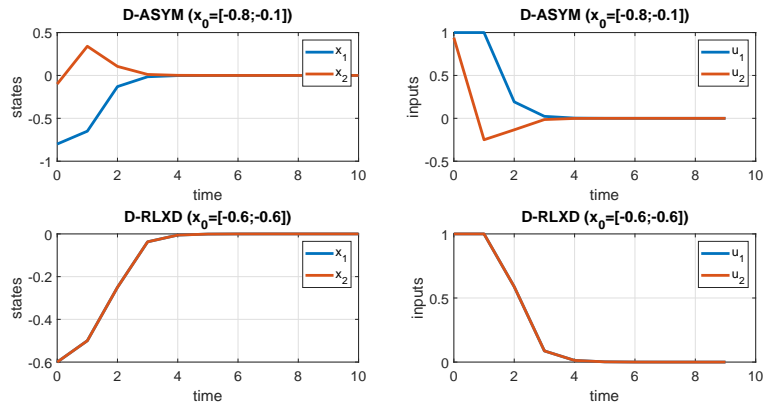}
	\caption{State and input trajectories of D-ASYM (with $x_0=[-0.8;-0.1]$) and D-RLXD (with $x_0=[-0.6;-0.6]$) when solving the optimization problem recursively}
\end{figure*}

\begin{table}
	\caption{Value of cost function for different schemes and initial conditions}
	\begin{tabular}{|c|c|c|c|}
		\hline
		Initial Conditions & 
		$x_0=\begin{bmatrix} -0.1 \\ -0.4 \end{bmatrix}$ & 
		$x_0=\begin{bmatrix} -0.8 \\ -0.1 \end{bmatrix}$ & 
		$x_0=\begin{bmatrix} -0.6 \\ -0.6 \end{bmatrix}$ \\
		\hline
		D-ADAP & 0.2528 & - & - \\
		\hline
		D-ASYM & 0.2528 & 1.5167 & - \\
		\hline
		D-RLXD & 0.2528 & 1.4192 & 1.8185 \\
		\hline
	\end{tabular}

\end{table}

\section{CONCLUSION}

In this paper, a novel distributed MPC scheme is developed with asymmetric adaptive ellipsoidal terminal sets. In this scheme, the size and the center of the terminal set is determined and updated online at each time instant taking into account the current state of the system. The positive invariance of the terminal set is ensured by imposing additional constraints in the MPC optimal control problem  on the size and the center of the terminal set. A relaxed version of this MPC scheme is developed by modifying the previously-added constraints. The proposed scheme and its relaxed version are compared to a recently-developed distributed MPC scheme and are found to be feasible even when this recently-developed scheme is not.

\bibliography{ifacconf}             

\begin{thebibliography}{19}
\providecommand{\natexlab}[1]{#1}
\providecommand{\url}[1]{\texttt{#1}}
\providecommand{\urlprefix}{URL }
\expandafter\ifx\csname urlstyle\endcsname\relax
  \providecommand{\doi}[1]{doi:\discretionary{}{}{}#1}\else
  \providecommand{\doi}{doi:\discretionary{}{}{}\begingroup
  \urlstyle{rm}\Url}\fi

\bibitem[{Bemporad and Morari(1999)}]{bemporad1999robust}
Bemporad, A. and Morari, M. (1999).
\newblock Robust model predictive control: A survey.
\newblock In \emph{Robustness in identification and control}, 207--226.
  Springer.

\bibitem[{Boyd et~al.(1994)Boyd, El~Ghaoui, Feron, and
  Balakrishnan}]{boyd1994linear}
Boyd, S., El~Ghaoui, L., Feron, E., and Balakrishnan, V. (1994).
\newblock \emph{Linear matrix inequalities in system and control theory},
  volume~15.
\newblock Siam.

\bibitem[{Boyd et~al.(2011)Boyd, Parikh, Chu, Peleato, Eckstein
  et~al.}]{boyd2011distributed}
Boyd, S., Parikh, N., Chu, E., Peleato, B., Eckstein, J., et~al. (2011).
\newblock Distributed optimization and statistical learning via the alternating
  direction method of multipliers.
\newblock \emph{Foundations and Trends{\textregistered} in Machine learning},
  3(1), 1--122.

\bibitem[{Christofides et~al.(2013)Christofides, Scattolini, de~la Pena, and
  Liu}]{christofides2013distributed}
Christofides, P.D., Scattolini, R., de~la Pena, D.M., and Liu, J. (2013).
\newblock Distributed model predictive control: A tutorial review and future
  research directions.
\newblock \emph{Computers \& Chemical Engineering}, 51, 21--41.

\bibitem[{Conte et~al.(2016)Conte, Jones, Morari, and
  Zeilinger}]{conte2016distributed}
Conte, C., Jones, C.N., Morari, M., and Zeilinger, M.N. (2016).
\newblock Distributed synthesis and stability of cooperative distributed model
  predictive control for linear systems.
\newblock \emph{Automatica}, 69, 117--125.

\bibitem[{Conte et~al.(2012)Conte, Voellmy, Zeilinger, Morari, and
  Jones}]{conte2012distributed}
Conte, C., Voellmy, N.R., Zeilinger, M.N., Morari, M., and Jones, C.N. (2012).
\newblock Distributed synthesis and control of constrained linear systems.
\newblock In \emph{2012 American Control Conference (ACC)}, 6017--6022. IEEE.

\bibitem[{Darivianakis et~al.(2019)Darivianakis, Eichler, and
  Lygeros}]{darivianakis2019distributed}
Darivianakis, G., Eichler, A., and Lygeros, J. (2019).
\newblock Distributed model predictive control for linear systems with adaptive
  terminal sets.
\newblock \emph{IEEE Transactions on Automatic Control}.

\bibitem[{Ellis et~al.(2014)Ellis, Durand, and
  Christofides}]{ellis2014tutorial}
Ellis, M., Durand, H., and Christofides, P.D. (2014).
\newblock A tutorial review of economic model predictive control methods.
\newblock \emph{Journal of Process Control}, 24(8), 1156--1178.

\bibitem[{Hovorka et~al.(2004)Hovorka, Canonico, Chassin, Haueter,
  Massi-Benedetti, Federici, Pieber, Schaller, Schaupp, Vering
  et~al.}]{hovorka2004nonlinear}
Hovorka, R., Canonico, V., Chassin, L.J., Haueter, U., Massi-Benedetti, M.,
  Federici, M.O., Pieber, T.R., Schaller, H.C., Schaupp, L., Vering, T., et~al.
  (2004).
\newblock Nonlinear model predictive control of glucose concentration in
  subjects with type 1 diabetes.
\newblock \emph{Physiological measurement}, 25(4), 905.

\bibitem[{Keerthi and Gilbert(1988)}]{keerthi1988optimal}
Keerthi, S.a. and Gilbert, E.G. (1988).
\newblock Optimal infinite-horizon feedback laws for a general class of
  constrained discrete-time systems: Stability and moving-horizon
  approximations.
\newblock \emph{Journal of optimization theory and applications}, 57(2),
  265--293.

\bibitem[{Klan{\v{c}}ar and {\v{S}}krjanc(2007)}]{klanvcar2007tracking}
Klan{\v{c}}ar, G. and {\v{S}}krjanc, I. (2007).
\newblock Tracking-error model-based predictive control for mobile robots in
  real time.
\newblock \emph{Robotics and autonomous systems}, 55(6), 460--469.

\bibitem[{Kouvaritakis and Cannon(2016)}]{kouvaritakis2016model}
Kouvaritakis, B. and Cannon, M. (2016).
\newblock Model predictive control.
\newblock \emph{Switzerland: Springer International Publishing}.

\bibitem[{Mayne et~al.(2000)Mayne, Rawlings, Rao, and
  Scokaert}]{mayne2000constrained}
Mayne, D.Q., Rawlings, J.B., Rao, C.V., and Scokaert, P.O. (2000).
\newblock Constrained model predictive control: Stability and optimality.
\newblock \emph{Automatica}, 36(6), 789--814.

\bibitem[{Mesbah(2016)}]{mesbah2016stochastic}
Mesbah, A. (2016).
\newblock Stochastic model predictive control: An overview and perspectives for
  future research.
\newblock \emph{IEEE Control Systems Magazine}, 36(6), 30--44.

\bibitem[{Prodan and Zio(2014)}]{prodan2014model}
Prodan, I. and Zio, E. (2014).
\newblock A model predictive control framework for reliable microgrid energy
  management.
\newblock \emph{International Journal of Electrical Power \& Energy Systems},
  61, 399--409.

\bibitem[{Rawlings and Muske(1993)}]{rawlings1993stability}
Rawlings, J.B. and Muske, K.R. (1993).
\newblock The stability of constrained receding horizon control.
\newblock \emph{IEEE transactions on automatic control}, 38(10), 1512--1516.

\bibitem[{Scherer et~al.(2014)Scherer, Pasamontes, Guzm{\'a}n, {\'A}lvarez,
  Camponogara, and Normey-Rico}]{scherer2014efficient}
Scherer, H.F., Pasamontes, M., Guzm{\'a}n, J.L., {\'A}lvarez, J., Camponogara,
  E., and Normey-Rico, J. (2014).
\newblock Efficient building energy management using distributed model
  predictive control.
\newblock \emph{Journal of Process Control}, 24(6), 740--749.

\bibitem[{Sznaier and Damborg(1987)}]{sznaier1987suboptimal}
Sznaier, M. and Damborg, M.J. (1987).
\newblock Suboptimal control of linear systems with state and control
  inequality constraints.
\newblock In \emph{26th IEEE Conference on Decision and Control}, volume~26,
  761--762. IEEE.

\bibitem[{Zeng and Wang(2015)}]{zeng2015parallel}
Zeng, X. and Wang, J. (2015).
\newblock A parallel hybrid electric vehicle energy management strategy using
  stochastic model predictive control with road grade preview.
\newblock \emph{IEEE Transactions on Control Systems Technology}, 23(6),
  2416--2423.

\end{thebibliography}
                                                   







\end{document}